# Unpacking the link between cousin marriage and women's paid work

**Sana Khalil**

**University of Washington (Tacoma)**

***Abstract***:

The debate surrounding the role of cousin marriage in women's autonomy, household status, and labor supply is longstanding and marked by contradictory viewpoints. Some studies suggest that cousin marriage enhances women's situation in the household, while others argue it restricts their freedoms and economic prospects. Despite this ongoing debate, quantitative investigations are limited. This study uses a sample of 15,068 married women from the Pakistan Demographic and Health Survey 2017-18 to examine the link between cousin marriage and women's labor supply patterns. The findings suggest a modest correlation between cousin marriage and reduced paid work. However, cousin marriage appears to have a more pronounced connection with women's work at home, potentially channeling them toward unpaid work for kin. Women in cousin marriages are unlikely to experience improved status within the household compared to women in non-cousin marriages. They are also more likely to rationalize acts of spousal violence in favor of patriarchal familial roles. In this regard, cousin marriage could potentially perpetuate patriarchal gender roles by penalizing women who deviate from conventional norms.

## I.  INTRODUCTION

Although marriages between blood relatives have declined in North America and Western Europe over the past century, they remain widely prevalent in the Middle East and most parts of South Asia. Bittles (2012) estimated that in Muslim countries of North Africa, Central and West Asia, and much of South Asia, marriages between second cousins or closer were around 20 to 50 percent. Consanguinity rates in the Muslim world may range between 20-60 percent (Hamamy, 2012). Pakistan has one of the highest rates of consanguineous marriages (marriages between blood relatives) in the world (Jones, 2010; Olubunmin et al., 2019). Consanguineous marriages[1]not only remain widely prevalent but have remained stable from 1990 to 2018. According to the Pakistan Demographic Health Survey (DHS) 1990-91[2], nearly 60 percent of marriages were consanguineous, one of the highest rates reported in the world (Hussain and Bittles, 1998). Of these consanguineous marriages, first-cousin marriages on the father's side were more common (nearly 30 percent). First-cousin marriages on the mother's side were relatively less common (21 percent). These patterns have shown little change over almost three decades. The Pakistan Demographic Health Survey 2017-2018[3] shows that nearly 64 percent of marriages in Pakistan are consanguineous, and almost half of all marriages nationally are between first cousins. Recent surveys also indicate the persistence of these marriage patterns in Pakistan, where first-cousin marriages on the father's side (29%) are more prevalent than those on the mother's side (21%). The prevalence and persistence of cousin marriages in Pakistan are worthy of attention, particularly because differences in marriage dynamics and prevailing gender norms can significantly influence socioeconomic outcomes and labor supply patterns for men and women (Dyson and Moore, 1983; Lenze and Klasen, 2017; Chiappori, 2020; Xiao and Asadullah, 2020; Calvo, Lindenlaub, and Reynoso, 2024). Consanguinity, which sustains the transmission of cultural values, strengthens social ties between family members, and maintains patrilineal kinship rules (Sandridge et al., 2010; Joshi, Iyer, and Do, 2014), may perpetuate patriarchal gender norms that influence women's labor supply patterns. In Islamic societies, where Islamic law mandates female inheritance, consanguinity can act as an informal institution to maintain consolidated land ownership, as marrying daughters to outsiders results in their shares of land leaving the male lineage (Mobarak et al., 2019; Brahmi-Rad, 2021). Consolidated land and kinship ties result in centralized decision-making, where the collective interests of the extended family (kin) often supersede those of the nuclear family. Consequently, male relatives can exert substantial control over women's labor, directing it in ways that maximize the interests of the group. While the literature has extensively examined the costs and benefits of cousin marriage (Joshi et al., 2014; Agha, 2016), its effect on women's paid work is relatively understudied.

---

[1] Excluding uncle-niece marriages which come under consanguineous marriages but are considered illegitimate in Pakistan.
[2] PHDS report 1990-91 noted that national-level data on consanguineous marriages had not been previously available for Pakistan. Therefore, for a comparative trend analysis in marriage patterns in Pakistan, the PDHS data from 1990-91 can serve as a reference point.

This study investigates the association between cousin marriage and women's household status, attitudes, and participation in paid work. Using data from the Pakistan Demographic Health Survey (PDHS) 2017-2018 (conducted by the National Institute of Population Studies and ICF) for 15,068 ever-married women ages 15-49, it analyzes the likelihood of women being in paid work[4] (defined as any work for which they receive cash payments) among those who marry their first or second cousins (a subset of consanguineous marriages) compared to those who do not. It also compares both groups of women across various indicators of intra-household bargaining and autonomy, e.g., participation in household decision-making, control over earnings, freedom of mobility, etc.

Research on consanguinity and cousin marriage suggests varied impacts on women's well-being. For example, women in cousin marriages may experience lower divorce rates[5] (Lenze and Klasen, 2017; Agha, 2016; Hamamy and Alwan, 2016; Charsley, 2007), reduced dowry costs[6] (Do et al., 2011; Anukriti and Dasgupta, 2017), and a higher degree of familiarity between bride/groom and in-laws (Sandridge et al., 2010). Additionally, they may benefit from greater financial stability, as familial unity ensures financial support in times of hardship (Weinreb, 2008). However, cousin

---

[3] The Pakistan Demographic Health Survey 2017-18 remains the latest demographic health survey so far.
[4] In PDHS 2017-18, around 15.35 percent of the married women report being in paid work while the rest don't work. Only 3.4 percent of married women (512 respondents out of 15068) report being in self-employment leaving a very small subsample to explore the effects of cousin marriage on other types of work and occupations.
[5] It is important to note that lower divorce rates may not always be beneficial for women as they may also suggest higher barriers for women to exit marital unions.
[6] Transfer of property, money, or other wealth by the bride's parents at the time of marriage (for a detailed discussion on the dynamics of dowry in Pakistan, see Makino, 2019)

marriages also present disadvantages and risks, such as an increased likelihood of recessive disorders in offspring (Tadmouri et al., 2009; Hamamy et al., 2012; Kelmemi et al., 2015), higher incidence of spousal violence[7] (Mobarak et al., 2019), and the reinforcement of patriarchal norms that limit women's autonomy and decision-making power (Charsley, 2007; Makino, 2019). Moreover, women in cousin marriages are more likely to marry early and start childbearing (Agha, 2016; Islam et al., 2018; Fatima and Leghari, 2020), which may restrict their opportunities for paid work and economic prospects outside the home.

More specifically, the literature offers contradictory hypotheses on the relationship between cousin marriage and women's paid work. Some argue that such marriages enhance women's autonomy and reduce constraints on their movements, potentially increasing their engagement in paid work due to strong kinship ties and stable family relationships. Dyson and Moore (1983), for example, argue that women in kinship endogamy (marriage within one's community, clan, or social group) in southern regions of India[8] enjoy greater autonomy and are more likely to pursue employment outside the home, compared to those in exogamous marriages in northern India. They attribute this to closer interactions with natal kin which facilitate familiarity with common rituals. As women in these types of marriages interact more actively with their natal kin than their counterparts (the closeness between the natal and affinal homes makes it easier for them to learn about the common rituals), their in-laws feel little need to 'repress', 'resocialize' or 'control' them to enforce social rituals. Consanguineous marriages diminish gender as a dominant factor in communication networks, enabling both men and women to freely socialize within their kinship circles without perceiving men from the same family as a threat (particularly in societies where strict gender segregation is enforced outside family contexts). This reduction in perceived threat may alleviate constraints on women's movements, thereby facilitating their pursuit of paid work. From this perspective, one might anticipate a positive effect of kinship endogamy and, by extension, of consanguineous or cousin marriages, on women's participation in paid work.

In contrast, others view cousin marriage as a patriarchal institution that reinforces strict male control over women's productive and reproductive capacities, favoring restrictive gender norms. Edlund (2018), for example, views cousin marriage as a symbol of male authoritarian power that constrains female autonomy and, consequently, economic development. Cousin marriage, Edlund argues, perpetuates predetermined unions and strengthens clan loyalty, contributing to cultural conservatism and societal stagnation. From this perspective, cousin marriages afford men greater control over marriage arrangements, limiting women's options for marriage and opportunities outside the home. Consequently, cousin marriage may be negatively associated with women's educational attainment and paid work (Agha, 2016; Khalil, 2021). Recently, Gevrek and Gevrek (2023) have shown that high rates of consanguineous marriage significantly predict a small, unexplained portion of the gender gap in labor force participation rates in Turkey, favoring males.

This stark contrast, in terms of how we understand the dynamics of cousin marriage and its implications for women's paid work, sets the context for this study. The purpose of this paper is not to settle this debate, per se; rather, it is an exploratory attempt to unpack the empirical linkages between cousin marriage and women's labor supply patterns as quantitative investigations on this subject are few and far between. It also examines how cousin marriage women's status in the household and their attitudes towards traditional gender norms, as proxied by a justification of spousal violence when a wife deviates from conventional gender roles.

---

[7] In the literature the terms intimate partner violence, domestic violence, and spousal violence are often used (see Vyas and Watts, 2009; Saud et al., 2021; Batthacharya, 2015). In this study, the term spousal violence is used to refer to the violence or abuse occurring between spouses who are legally married to each other, particularly considering that in Pakistan, cohabitation of opposite sexes without marriage is illegal. The evidence on the link between consanguineous marriage and intimate partner violence is mixed (see Hamamy and Alwan, 2016).

[8] Uncle-niece marriages are more prevalent in southern regions of India compared to northern regions of India where such unions are strictly prohibited.

Theoretically, I identify two broad channels, among others, through which cousin marriage might influence married women's participation in paid work. Firstly, since women who marry their cousins are more likely to reside with their in-laws (in an extended family system), given that cousin marriage is a form of consolidation of kinship ties (Fatima and Leghari, 2020), they might face a larger burden of unpaid domestic care than women who marry outside the family.[9] This would lower consanguine women's participation in paid work. Secondly, women in cousin marriages, as newly added members of extended family, experience pressures to conform to a standardized model of behavior that aligns with the expectations of their affinal family. This model often entails domestic labor for the affinal family(Akram-Lodhi, 1996; Weinreb, 2008). Additionally, these women may face heightened scrutiny and surveillance, particularly if their natal and affinal homes are close, to ensure their adherence to societal ideals of "purity," which often entails limiting interactions with unrelated men (Jayachandran, 2021). Moreover, in Pakistani society, women are commonly regarded as secondary earners, with a higher likelihood of entering the workforce to compensate for any financial shortfalls within the family or to provide additional support (Khan and Khan, 2009). However, the decision for women to engage in paid work can be socially construed as a sign of a strained domestic financial situation, potentially diminishing the household's social standing. Women in cousin marriages may bear a heavier burden of expectations to remain at home and contribute towards household production (offering unpaid labor to meet domestic needs) and prestige (adhering to social ideals of 'femininity').

Finally, cousin marriages may afford women various resources embedded in kinship structures (see Joshi et al., 2014; Agha, 2016). The lower threat of divorce in cousin marriages and the immediate availability of a social network, including elderly family members who can arbitrate conflicts (see Charsley, 2007), suggest that kinship affinity and group loyalty could mitigate some of the negative aspects associated with cousin marriage. Moreover, women may rely on their kin to care for their children while they engage in paid work outside the home. Therefore, one might think of a positive relationship between cousin marriage and women's paid work, similar to the position upheld by Dyson and Moore. However, it is important to note that cousin marriages may also entail challenges. Male relatives may exert greater control over women in these marriages, which may not necessarily improve their status or autonomy within the household.

Given the complex interactions of these factors, the overall effect of cousin marriage on women's paid work appears ambiguous. Using data from 15068 ever-married women from the Pakistan Demographic Health Survey (PDHS) 2017-18, I examine the following key propositions: 1) Cousin marriage is associated with reduced participation of women in paid work and increased involvement in work at home, 2) Cousin marriage may constrain women's position in the household and perpetuate existing gender norms, such as the justification of violence against a wife if she does not conform to her expected gender roles. This paper contributes to the literature by exploring the relationship between cousin marriage and women's labor supply patterns. It aims to identify how cousin marriage influences women's attitudes towards gender roles and various indicators of intra-household bargaining, which in turn affect women's educational attainment and participation in market labor. Findings from this paper suggest that while cousin marriage may have a weak negative association with women's engagement in paid work, it could have significant implications for the quality of their work, potentially leading to increased unpaid labor for kin. Additionally, to the best of my knowledge, this is the first study that quantitatively examines the link between cousin marriage and spousal violence, revealing that women in cousin marriages may be more likely to rationalize spousal acts of violence in favor of traditional gender roles. These findings contribute to our understanding of marriage dynamics, gender norms, and women's attitudes toward spousal violence in the developing world, particularly within Muslim-majority countries.

---

[9] In Pakistani society, daughters in laws generally face a disproportionately higher burden of unpaid domestic care than other women in the household, like sisters-in-law and mothers-in-law.

## II.  RELATED LITERATURE AND CONCEPTUAL LINKS

At 21 percent, Pakistan's female labor force participation rate is one of the lowest in both South Asia and globally. The majority of women workers are represented in the informal sector as contributing family workers, with 54% of women compared to 10% of men. A significant proportion of these women are unpaid (World Bank Data, 2022). Due to limitations in data collection, estimates likely underrepresent the true extent of women's paid and unpaid work. Previous literature from Pakistan has examined the effect of personal and family characteristics as the primary factors influencing women's labor (Shah 2004; Faridi and Rashid, 2014). However, the role of marriage and household dynamics remains unexplained.

This paper focuses on how marriages to first and second cousins (a subset of consanguineous marriages) affect women's likelihood of being in paid work. However, discussing consanguineous marriages in general is necessary to understand contextual complexities. Current evidence on cousin marriage and female labor force participation is limited, with a particular focus on regions characterized by high consanguinity rates such as the Middle East, North Africa, and South Asia. While most studies have focused on the correlates of consanguinity (Bittles, 2012; Do et al.; Mobarak et al., 2013; Akyol and Mocan, 2023), few have directly explored its impact on women's labor force participation. Brahmi-Rad (2021) found that female inheritance increases the likelihood of cousin marriage and reduces women's economic participation in India. Righetto's (2023) analysis of data from Italian municipalities showed that endogamous marriages widened the gender gap in labor force participation, reinforcing social norms stigmatizing women's participation in paid work.

Similarly, Bhopal et al. (2014) examined the social and economic impacts of consanguineous marriages among Pakistani-origin women in the UK, revealing lower employment rates and educational attainment among those in such marriages compared to non-consanguineous counterparts. In Pakistan, Agha (2016) conducted a qualitative study suggesting that women in consanguineous marriages are less likely to pursue education and tend to marry early. Despite the various disadvantages associated with cousin marriages, women often prefer them. Chaudhry and Arif (2023) surveyed nine districts of Pakistan and observed that restrictive gender norms linked with consanguinity can be transmitted intergenerationally. Their findings indicated that children of parents who are first cousins are less likely to receive education, with investments particularly lower for daughters.

While evidence on the link between cousin marriage and women's involvement in paid work is limited, various other studies shed light on its broader social and economic effects. In the Middle East and Asia, where first- and second-cousin marriages are widely prevalent (Hamamy et al., 2011), studies have found high rates of consanguineous marriages in rural areas where women are less educated, face restrictions on their mobility, and marry early (Shah, 2004; Koc, 2008;

Weinreb, 2008; Islam, 2018). In Kuwait, Shah (2004) notes that preserving the "purity" of lineage is an important motivating factor for marrying a biological relative. Similar convictions are widely prevalent in the Middle East, especially in Arab communities, that exogamous marriages weaken a family's bloodline and inheritance (Weinreb, 2008; Harkness and Khaled, 2014). In the South Asian context, a conspicuous feature of consanguineous marriage is lower dowery payments (Do et al., 2013; Mobarak, Kuhn, and Peters, 2013, Agha, 2016) as well as deferral of dowry payments until after marriage (Mobarak et al., 2019). Quantitative evidence from India and Pakistan also suggests that women in consanguineous marriages have lower education and socioeconomic standing and live in extended family systems (Hussain and Bittles, 1998; Sharma et al., 2020; Rahman et al., 2021).

Using the insights from the related literature, we can envisage three important channels, among others, through which cousin marriage influences married women's involvement in paid work. The first channel emanates from the role of cousin marriage in status production[10] and its role in reinforcing restrictive gender norms. Various studies have documented why consanguinity, especially marriages to first and second cousins, has persisted in the MENA region and various parts of Asia: preservation of social status, family honor, and lineage are among the key motivating factors (Jurdi and Saxena, 2003; Weinraub, 2008; Jacoby and Mansuri, 2010; Harkness and Khaled, 2014; Agha, 2016). Women in cousin marriages may face higher expectations of kin-based altruism given their familiarity and embeddedness in extended family systems. This can lead to greater scrutiny of their adherence to familial norms and customs, potentially limiting their mobility and opportunities for paid work outside the home. As insiders, women married to cousins are often perceived as custodians of household honor and may face more severe penalties for deviating from expected roles compared to women married to non-relatives—considered outsiders. This dynamic reflects the patriarchal and hierarchical nature of the institution of marriage in which alliances are formed to consolidate inter-group cohesion (Dyson and Moore, 1983).

Additionally, the role of women in status production, particularly in hosting large social gatherings, places a significant burden on women of kin. In certain societies like Pakistan, women are expected to personally attend to guests, leading to increased unpaid labor (see Akram-Lodhi, 1996). Consequently, women in cousin marriages may allocate more time to status production, potentially limiting their engagement in paid work or confining it to home-based informal employment. Similarly, cousin marriage may pose constraints on women in terms of a heavier burden of unpaid care work as couples in these marriages are more likely to live in an extended family system (Jurdi and Saxena, 2003; Kaplan et al., 2016; Khan, 2021) and have larger family size (Islam et al, 2018) than couples in exogamous marriages. Within extended family arrangements, daughters-in-law often bear a greater responsibility for unpaid domestic care, traditionally tasked with managing the household affairs while assistance from sisters-in-law and mothers-in-law may be limited (see Turaeva and Becker, 2022).

The second channel through which consanguinity influences married women's labor supply revolves around the preservation of property ownership and inheritance. Under Islamic inheritance law, women have a de jure claim over the property of their parents and husbands. Cousin marriages often lead to the consolidation of inherited wealth within the extended family, as husbands and consanguineous in-laws become de facto claimants to a woman's inheritance.[11] This reflects a patriarchal marriage dynamic that influences negotiations over women's property and future wealth. In consanguineous marriages, if a woman's inheritance is transferred to her husband or in-laws, her bargaining power within the household may diminish, leaving her in a weaker position during intra-household negotiations. In other words, the loss of ownership and control over assets may diminish women's bargaining power within the household (see Agarwal, 1994 p.53-62). This could result in a heavier burden of unpaid domestic work, limiting her ability to engage in paid employment. Conversely, if consanguinity results in such negotiations where a woman, due to the affinity (sustained via individual emotional pressures and social networks) with in-laws, can maintain de jure control of her inheritance, her bargaining position within the household may strengthen. This could lead to negotiations for reduced domestic responsibilities and increased support with childcare, potentially facilitating her participation in paid work, depending on her work preferences.

---

[10] Domestic work at home is often considered 'status-producing' for households and is associated with higher household income (Naidu, 2016). In the Indian context, according to Eswaran, Ramaswami, and Wadhwa (2013), avoiding work involving contact with non-family males is crucial for maintaining status. Additionally, activities labeled as 'status production' require married women's time and labor. The poorest individuals lack assets, typically land, relying solely on labor for income. However, as incomes rise, concerns about status become more prominent, leading married women to gradually withdraw from market work. Eventually, at higher levels of affluence, they may cease market work entirely to focus on status production.

A third channel can be envisaged in terms of differences in prevalent marriage rules— e.g. bride price and dowry[12]— associated with consanguineous marriages and out-marriages. Dowry and bride prices have been prevalent elsewhere. Dowry, historically practiced in medieval Europe and other regions, influences household chores allocation and decision-making authority (Anderson, 2003). In contrast, bride price, prevalent in sub-Saharan Africa and other societies (Anderson, 2007), reflects transactions between families and signifies duties related to labor and reproduction. Negotiations over dowry and bride price may differ based on social status, with consanguineous marriages potentially involving lower payments due to mutual agreements among kin. Dowry and bride prices significantly impact the bargaining positions of men and women within households, leading to varying outcomes across close-kin marriages and out-marriages. For instance, research in China by Brown (2009) showed that higher dowries lead to increased allocation of household chores by husbands. Dowry also positively affects the wife's leisure time and decision-making authority.

Negotiations between the families of bride and groom, over dowry and bride price, may vary based on the social status of both parties —if both parties are equal in status, no exchange might arise (see Agarwal, 1994, p. 126). In consanguineous marriages, the bride price and dowry might be lower due to mutual negotiations among related families, driven by factors such as group solidarity and kin-based altruism, as well as a lower perceived threat of divorce. The bride price can be viewed as a form of compensation for the daughter's labor power transferred to her husband and in-laws, particularly linked to her duty to bear children, termed as 'child price'- where the rights of children are transferred to the father's consanguine group (see Rudwick and Posel, 2013). This dynamic, mediated by consanguinity, can lead to complex outcomes in intra-household bargaining over the allocation of time and resources among the wife, husband, and in-laws. Intra-household bargaining in this context can be understood better under Folbre's (2006) conceptual framework on collective conflict and gender roles in bargaining between men and women under patriarchal systems. Folbre notes that fallbacks (and hence bargaining outcomes) of men and women are determined, in part, by social institutions that dictate marriage rules. Initial differences in endowments and resources can result in bargaining outcomes more advantageous to men. In this context, consanguinity can influence intra-household bargaining by mediating dowry and bride price negotiations. However, it remains unclear whether this strengthens or weakens women's bargaining power.

---

[11] Brahmi-Rad (2021) provides evidence from Indian National Family Health Surveys that women's inheritance rights in Islam contribute to the prevalence of cousin marriage in Muslim communities to preserve their share in the male line and pool family assets within the patrilineal lineage, as daughters contesting for their share of inheritance could jeopardize family assets and social cohesion.

[12] Dowry is the transfer of goods or money from the bride's family to the groom or his family, ensuring the bride's financial security or demonstrating her social status. On the other hand, bride price involves the payment from the groom or his family to the bride's family as compensation for the marriage, typically recognizing and compensating for the loss of the daughter's labor and companionship (Anderson, 2003).

It is also important to note that consanguineous marriage offers male relatives opportunities to bypass customary payments to women such as bride price, or reduce the amount of mandatory payments like *mahr*[13] (Edlund, 2018). Due to their familial affinity, grooms and their families often negotiate marriage contracts with a better position, potentially lowering the bride price and *mahr*. In these negotiations, male kin, acting as the bride's guardian (*wali*), typically negotiate with the groom's appointed *wali*, while the bride is usually not directly involved. Cousin marriage's elimination or reduction of bride price, dowry, and *mahr* can limit women's bargaining power over domestic chores and external opportunities.

The mechanisms, identified here, carry both negative and positive implications for women's participation in paid work, leaving the overall effect ambiguous. Cousin marriage can reduce women's labor supply by reinforcing social norms that stigmatize working women, increasing costs for deviating from traditional gender roles, and perpetuating restrictive gender norms—a gap in the literature that this study aims to address. Specifically, this study will test the following hypotheses:

1) Cousin marriage is associated with reduced participation of women in paid employment in Pakistan.
2) Cousin marriage may constrain women's position in the household and perpetuate existing gender norms, such as the justification of violence against a wife if she does not conform to her expected gender roles
3) The incidence of cousin marriage encourages women's involvement in domestic labor at home

## III. DATA AND DESCRIPTIVE STATISTICS

Data for this study comes from the Pakistan Demographic Health Survey (PDHS) 2017-18, conducted jointly by the Pakistan Bureau of Statistics, the National Institute of Population Studies, and USAID.[14] The dataset comprises information gathered from 15,068 ever-married women aged 15-49 years, primarily drawn from this survey unless specified otherwise. The survey covers a wide range of variables including personal and household characteristics, participation in paid work (either for cash or in-kind), blood relation with husband, involvement in family decision-making (such as control over personal health decisions, household purchases, earnings, and decisions regarding family visits), and attitudes towards justification of violence against a wife in various scenarios (such as going out without telling her husband, neglecting children, arguing with her husband, burning the food, and refusing to have intercourse with her husband), among other relevant areas. Table 1 presents the descriptive statistics for the key variables in this study.

---

[13] *Mahr*, an Arabic term meaning 'gift,' refers to the payment that a groom is obligated to pay the bride at the time of marriage under Islamic law. Part of this payment can be deferred and paid in the event of marriage dissolution. Unlike bride price, *mahr* is de jure obligatory and can only be waived under the wife's free will, as per the Islamic contract of marriage (*Nikah*). *Mahr* is often stipulated in the Islamic marriage contract and its payment is enforceable under Islamic law as it imposes a financial obligation on the husband (Qaisi, 2001). In contrast, bride price and dowry are optional customary practices and are not mandatory under Islamic law.

[14] DHS 2017-18 is the fourth survey that was conducted in Pakistan. The first demographic health survey was conducted in 1990-91. The second DHS was published in 2006-07, while the third one was conducted in 2012-13.

**Table 1: Summary statistics**

| | N | Mean | SD | Min | Max |
|---|---|---|---|---|---|
| **Work status:** | | | | | |
| Paid work (1=work for cash or in-kind remuneration, 0 if not) | 15060 | 0.156 | 0.363 | 0 | 1 |
| Work at home[15] (1= work at home, 0=away) | 2320 | 0.486 | 0.500 | 0 | 1 |
| Occupation type (1=service-related, 0= agricultural or manual) | 2312 | 0.384 | 0.486 | 0 | 1 |
| **Type of marriage** | | | | | |
| Cousin Marriage(1=husband is a first- or second-cousin, 0=distant or no blood relation with husband) | 15068 | 0.558 | 0.497 | 0 | 1 |
| **Personal and household characteristics:** | | | | | |
| Respondent's age | 15068 | 32.319 | 8.324 | 15 | 49 |
| Respondent's education | 15068 | 4.436 | 5.234 | 0 | 16 |
| Education of Husband | 14469 | 7.008 | 5.274 | 0 | 16 |
| Husband's work status | 15068 | 0.946 | 0.226 | 0 | 1 |
| Number of Children | 15068 | 1.451 | 1.525 | 0 | 13 |
| Gender of household head (1=male, 0=female) | 15068 | 0.890 | 0.313 | 0 | 1 |
| Wealth index[16] (divided into five quintiles: poorest, poorer, middle, richer, and richest) | 15068 | 3.004 | 1.412 | 1 | 5 |
| Ethnicity (categories: English, Urdu, Sindhi, Punjabi, Saraiki, Baluchi, Pushto and others) | 15068 | 5.646 | 2.040 | 2 | 8 |
| Rural (1= rural area, 0= urban) | 15068 | 0.519 | 0.500 | 0 | 1 |
| Region (Punjab, Sindh, Khyber Pakhtunkhwa, Balochistan, Gilgit Baltistan, Islamabad, and two federally administered and controlled areas) | 15068 | 3.626 | 2.289 | 1 | 8 |
| **Participation in household decision-making[17] (1= self or jointly with husband, 0= husband alone or someone else):** | | | | | |
| Decision to visit family/relatives | 15068 | 0.447 | 0.497 | 0 | 1 |
| Decisions about self-earnings | 15068 | 0.111 | 0.314 | 0 | 1 |
| Decisions about household purchases | 15068 | 0.399 | 0.490 | 0 | 1 |
| Decisions about personal health care | 15068 | 0.464 | 0.499 | 0 | 1 |

[15] Work at home includes paid and unpaid work for relatives (e.g., working on a relative's farm) whereas work away from home denotes paid work outside the household.

[16] In DHS survey data, wealth index is calculated based on a principal component analysis based on household ownership of assets ranging from televisions, bicycles, etc. to materials used for housing construction as well as access to water and sanitation facilities. For more details see (https://www.dhsprogram.com/pubs/pdf/CR6/CR6.pdf)

[17] Each decision-making variable has the following categories: 1) Respondent alone makes the decision, 2 ) Respondent and husband/partner jointly make the decision, 3) Husband/partner alone makes the decision, 4) Someone else other than the respondent or their husband/partner makes the decision, and 5)Other (not specified). Given the small subsamples associated with each category and the study objective to examine women's involvement in the decision-making process, each variable is recoded as follows: 1 if a woman was consulted alone or jointly with her husband in the decision-making process, and 0 if she was not involved or if her husband or someone else was involved.

| Women's attitudes toward spousal violence[18] (1= yes; 0=no): Is it justified for a husband to hit his wife when she: | | | | |
|---|---|---|---|---|
| Goes out without telling husband | 14743 | 0.371 | 0.483 | 0 | 1 |
| Neglects children | 14709 | 0.308 | 0.462 | 0 | 1 |
| Argues with her husband | 14753 | 0.286 | 0.452 | 0 | 1 |
| Refuses to have intercourse with husband | 14524 | 0.316 | 0.465 | 0 | 1 |
| Burns the food | 14681 | 0.194 | 0.395 | 0 | 1 |

**The outcome variables:** Regarding women's participation in paid work, there are three outcome variables of interest: women's paid work for cash or in-kind remuneration after marriage, work at home, and type of occupation. The primary explanatory variable of interest, paid work, takes a value of 1 if a woman works for cash or in-kind remuneration, and 0 otherwise. To test the hypothesis in this study that cousin marriage may impose various constraints on women, including a heavier burden of unpaid domestic care, the second primary variable of interest is work at home (equal to 1 if a woman works at home, 0 if she works outside the household). This includes informal employment in the form of contributing family work.[19] The third outcome variable of interest is the type of occupation. According to PDHS 2017-18, out of 15068 ever-married women, nearly 12748 (85 percent) are not currently working, 512 are in agricultural self-employment (3.4 percent), 791 are in skilled manual labor (5.25 percent), 512 are in professional/technical/managerial work (3.4 percent), 293 work in other services (1.95 percent), and only 121 (0.80 percent) are in unskilled manual work. The limited size of these occupational categories poses challenges in exploring how cousin marriage may influence women's occupational choices. Consequently, for analytical purposes, I categorize occupations into two broad groups: service-related occupations (coded as 1) and agricultural and manual work (coded as 0). Overall, approximately 888 respondents are employed in service-related occupations, while 1,424 are engaged in agricultural and manual occupations. The objective is to analyze whether being in a cousin marriage is associated with differences in married women's choices of occupations. Moreover, to examine how cousin marriage influences women's status in the household and attitudes towards conventional gender roles and norms, variables on household decision-making and spousal violence are included.

---

[18] The term 'spousal violence' specifies violence or abuse occurring between spouses who are legally married to each other, particularly considering that in Pakistan, cohabitation of opposite sexes without marriage is illegal. This term specifically addresses the violence experienced by married women at the hands of their husbands, which is the focus of this study. While 'spousal violence' falls under the broader category of 'domestic violence,' it pertains specifically to violence within the context of marriage. On the other hand, 'domestic violence' encompasses a wider range of abusive behaviors within familial and household relationships. This includes violence against children, elderly family members, or other household members, indicating that it can involve violence experienced by married women from husbands, in-laws, and other relatives. Domestic violence is not limited to marital relationships and can occur in various types of relationships. In Pakistan Demographic Health Surveys, the term 'wife-beating' has been used to capture spousal violence. This term highlights cultural and religious attitudes and norms that normalize or justify violence against a wife by her husband.

[19] According to International Labor Organization (ILO) definition, contributing family workers are individuals who work in self-employment roles within market-oriented establishments operated by a relative residing in the same household.

**The independent variables:** The primary explanatory variable in this study is cousin marriage which narrows down the blood relation in terms of marriage to a first-and second-cousin (coded as 1/0). This variable captures the closeness between affinal and natal families, as discussed conceptually above. For comparison, I also examine the role of consanguinity, or general blood relation with the husband (coded as 1 if the husband is a woman's first cousin from either the paternal or maternal side, or a second cousin, or another distant blood relative, and 0 if there is no blood relation).

In multivariate regressions, several control variables are included to mitigate the risk of spurious relationships and omitted variable bias. These controls include the respondent's age, level of education (years of schooling), husband's education (years of schooling), husband's employment status, number of children, gender of the household head, wealth status of the household (in terms of 5 quintiles: poor, poorer, middle, rich, and richer), and indicators for ethnicity and rural/urban residence. The rationale behind including these control variables lies in their documented effects on labor market outcomes, as supported by existing literature (Naqvi, Shahnaz, and Arif, 2002; Khalil, 2021). Previous research suggests that factors such as women's age, education, household poverty, and husband's unemployment positively and significantly affect female labor force participation. Conversely, the presence of children, household assets, and the husband's education are negatively related to women's participation in paid work (Khan and Khan, 2009; Faridi and Rashid, 2014). Similar control variables have been highlighted in other studies on women's employment. For example, Klasen and Pieters (2012) found that in India, less-educated women were often "pushed" into the labor force by economic necessities such as their husband's unemployment, lower household income, and reduced social status. In contrast, highly educated women were "pulled" into the labor market by the prospect of attractive employment opportunities and higher wages.

To examine women's status in the household, additional indicators related to their involvement in household decision-making, freedom of movement, and attitudes toward violence against wives are explored (as detailed in Table 1). These indicators have been widely used in studies to explore women's instrumental agency and bargaining power within the household (Yount et al., 2016; Yount et al., 2019). Scholars have linked women's autonomy and empowerment to their ability to make independent decisions (see Malhotra and Schuler, 2005; Sraboni et al., 2014; Hanmer and Klugman, 2016; Miedema et al., 2018). However, there exists a rich debate surrounding the choice of indicators for measuring women's empowerment (Chowdhury et al., 2018).

The Pakistan Demographic and Health Survey (PDHS) 2017-18 provides limited indicators of women's participation in household decision-making: decisions about visits to family or relatives, women's health care, major household purchases, spending husband's earnings, and own earnings. For this study, these indicators are coded as 1 if a woman makes these decisions individually or jointly with her husband, and 0 if the decisions are made solely by the husband or someone else.

Similarly, acceptance of spousal violence or "wife-beating" (terms often used interchangeably with domestic violence or intimate partner violence) reflects attitudes about women's general roles in the household and society (Haj-Yahia, 1998, 2003; Boy and Kulczycki, 2008; Zaatut and Haj-Yahia, 2016). Women's approval/justification of spousal violence, for specific actions and situations, has been ascribed to their acceptance and internalization of patriarchal social norms (Ahmad et al., 2004; Dhanaraj and Mahambare, 2022) and constrained individual agency (Hanmer and Klugman, 2016). There exists a rich literature exploring reasons behind women's acceptance/justification of spousal violence (see Dasgupta, 2019 for key propositions). As discussed earlier, in the Pakistani context, women in cousin marriages may face stricter adherence to traditional gender roles and norms. This perspective suggests a potential positive association between cousin marriage and women's justification of spousal violence under certain circumstances.

Table 1 provides descriptive statistics for these variables. A few key features are discussed here. Out of 15068 ever-married women, only 2347 (15.6%) report being in paid work, with the majority engaged in work at home. Nearly 62% of women report having a blood relation with their husbands before marriage. In the sample, women on average have 4.4 years of education compared to 7 years of their husband's education. Nearly 89% of the households are headed by men (only 11% are headed by women). Women's participation in household decision-making is highly restricted, especially when it comes to decisions about their earnings: only 11% report involvement (alone or jointly with husbands) in decisions about controlling their earrings. Similarly, 55% of the women have no say in planning visits to their family and relatives indicating restrictions in their mobility. In terms of attitudes towards justification of spousal violence, a substantial number of women report that a husband is justified in hitting the wife if she goes out without telling her husband (37%), neglects the children (31%), argues with her husband (29%), refuses to have intercourse with her husband (32%), and burns the food 19%).

Table 2 examines the proportion of women in cousin and non-cousin marriages who had a say in choosing their husbands. Say in choosing a husband means that the survey participant indicated that she had a choice when selecting a husband[20] (though it is challenging to ascertain whether the marital decision wasn't influenced by indirect expectations). Approximately 83% of women in cousin marriages and nearly 77% in other marriages report involvement in selecting their husbands (Chi-square test, $p < 0.001$). This is consistent with existing literature suggesting that consanguineous marriages may be preferred by women due to the combination of costs and benefits associated with them (Agha, 2016; Mobarak et al., 2019). Nonconformity to prevailing gender norms imposes significant costs and risks on women (Eswaran et al., 2013; Jayachandran, 2021), which may potentially motivate them to opt for marriage within their extended family despite its disadvantages. Additionally, when choosing a spouse, a strong sense of familiarity, trust, and shared values between families may play a significant role.

**Table 2: Cousin marriage and say in choosing a husband**

| Cousin Marriage | Say in Choosing Husband | | |
|---|---|---|---|
| | No N (%) | Yes N (%) | Total |
| Other marriages | 1511 (23) | 5127 (77) | 6638 |
| Cousin marriage | 1434 (17) | 6949 (83) | 8383 |

---

[20]In the Pakistan Demographic Health Survey 2017-18, a woman's consent in marriage is represented by a dichotomous variable named "had a say in choosing the husband," where 1 indicates "yes" and 0 indicates "no."

Figure 1 shows that among women who have a blood relationship with their husbands, marriages to first cousins on the father's side are more prevalent (45%) compared to marriages to first cousins on the mother's side (30%), second cousins (14%), or other relatives (10%). This trend has been consistently observed in Pakistan, with first-cousin marriages from the father's side being more prevalent (Hussain and Bittles, 1998; Agha, 2016). Similar patterns are also observed in India and Bangladesh, where paternal first-cousin marriages are preferred among land-owning classes to consolidate land parcels within the paternal lineage (Mobarak et al., 2019; Brahmi-Rad, 2021).

**Figure 1: Incidence of cousin marriage in Pakistan, 2017-18**

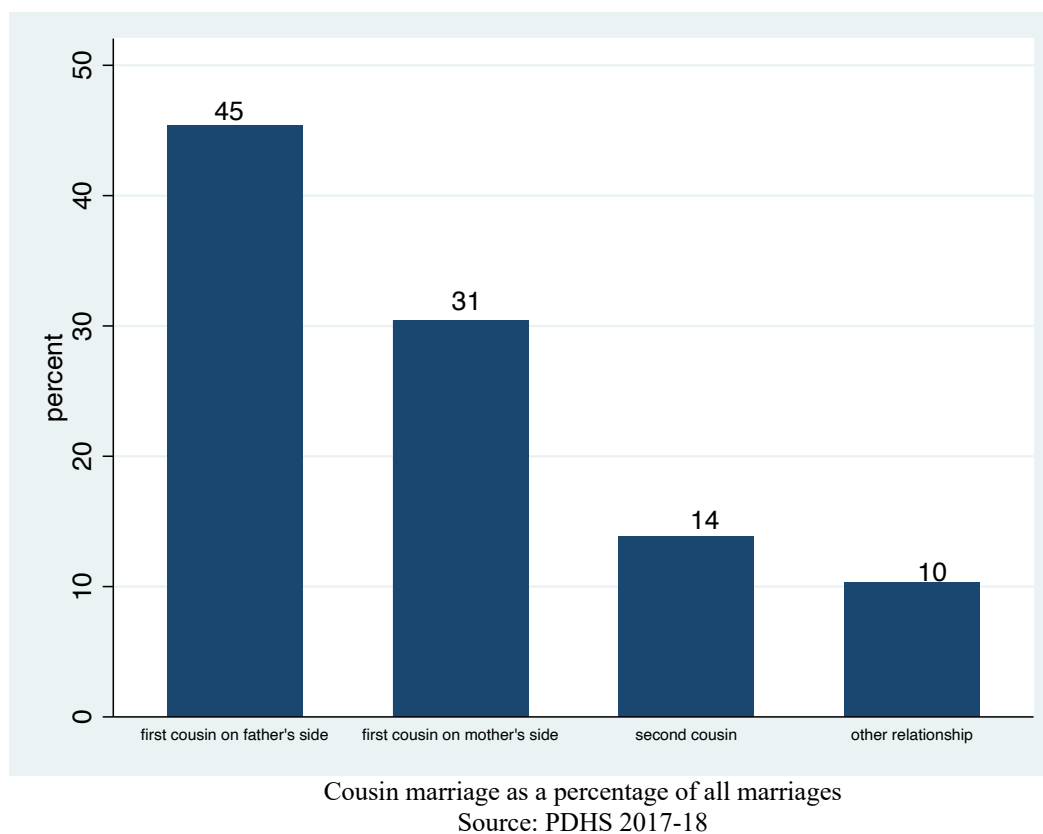

Cousin marriage as a percentage of all marriages
Source: PDHS 2017-18

The next section explores the relationship between cousin marriage and married women's labor supply patterns, as well as their status within the household and attitudes toward spousal violence.

## IV. Empirical estimation and results

### a) Descriptive analysis: Cousin marriage, mobility, and participation in major household decisions

I begin by testing hypotheses regarding how cousin marriage may influence women's status within the household and their attitudes toward traditional gender roles expected of a wife. As discussed previously, women's ability to participate in household decision-making may indicate their general status in the household. While women's empowerment is a multidimensional concept (Kabeer 2011), it includes their ability to make choices, including decisions about their movement and well-being, and expression of equitable gender beliefs (Miedema et al., 2018). In Pakistan's context, Abbas et al. (2021) use data from PDHS 2012-13 and PDHS 2017-18 to show that number of children, residence in urban areas, and participation in paid work (both skilled and unskilled) are positively associated with women's greater involvement in household decision-making in Pakistan. There is substantial literature that uses indicators on women's participation in household decision-making from various Demographic Health Surveys to capture women's bargaining power and autonomy in the household (see Hanmer and Klugman, 2016; Miedema et al., 2018; Chowdhury et al., 2018; Dhanaraj and Mahambare, 2022), however, little is known how marriage dynamics can influence women's status in the household. Table 3 uses a chi-square ($\chi^2$) test of proportions to examine how cousin marriage is associated with each indicator of women's participation in household decision-making.

Table 3 indicates that, for most women regardless of marriage type, decisions about family visits, which often reflect freedom of mobility, are primarily made by husbands or other family members.

**Table 3: Participation in household decision-making by the type of marriage**

|  | Women in non-cousin marriages (n=6,656) | Women in cousin marriages (n=8412) | P-value |
|---|---|---|---|
| Decision to visit family/relatives | 3046 (46) | 3687 (44) | 0.018 |
| Participant's healthcare | 3172 (48) | 3821 (45) | 0.006 |
| Major household purchases | 2758 (41) | 3254 (38) | 0.001 |
| Decisions about husband's earnings | 2779 (42) | 3383 (40) | 0.055 |
| Decisions about own earnings | 760 (11) | 914 (10.8) | 0.283 |

Notes: Each of the decision-making variables[21] is coded as 1 if a woman alone or jointly with her husband is consulted in the decision-making process; 0 if she is not involved in the decisions (husband alone or someone else is involved). The table shows the number of respondents, with percentages in parentheses, who reported participation in these decisions alone or jointly with their husbands. The last column shows the p-values for a test of proportion (under the Chi-Square test) with the null hypothesis of no difference across groups.

---

[21] These variables capture who typically makes decisions regarding different aspects of women's intrinsic agency. In PDHS survey each decision variable has the following categories: 1) Respondent alone; 2) Respondent and husband/partner; 3) Husband/partner alone; 4) Someone else (can be another relative or person); 5) Other (represents other individuals who make the decision but do not fit into the previous categories). For an ease of analysis, table 3 presents the binary recoding of the variables; 1= participation in the decision alone or jointly with husband, 0= no participation in the decision-making process.

Notably, only 11 percent of women in both cousin and non-cousin marriages report being involved in decisions about their own earnings, either alone or jointly with their husbands. In the majority of cases, these decisions are made solely by the husband or another family member. Women in non-cousin marriages appear more likely to participate in decisions about major household purchases compared to those in cousin marriages. The p-value of 0.001 shows a highly significant difference in this area. Further analysis of these associations is presented in Table A in the appendix.

Table A presents the results of a logit regression model examining the relationship between cousin marriage and various aspects of women's status in the household while controlling for other relevant factors. The findings from Table A indicate that cousin marriage is not significantly associated with decisions related to visits to the family, healthcare, or decisions about husbands' earnings. However, the analysis reveals a statistically significant negative association between cousin marriage and women's involvement in decisions regarding household expenditures and personal income. This could be due to cultural or familial dynamics associated with cousin marriages that limit women's influence over financial matters. Several other noteworthy findings emerge from Table A. Women's age is positively associated with women's involvement in various household decision-making processes, indicating a trend towards greater influence with age. This may be attributed to older women being perceived as more knowledgeable or authoritative.

Moreover, women with higher levels of education are more likely to participate in decision-making across all aspects. However, when the household head is male (coded as 1), women are significantly less likely to be involved in decisions related to visits to family, healthcare, purchases, and husband's earnings compared to when the household head is female (coded as 0). This highlights a gender disparity in decision-making autonomy within households, where male household heads may exert more influence over such decisions.

Overall, the results suggest that cousin marriage does not improve women's status through increased participation in household decision-making. In fact, it appears to be negatively associated with women's likelihood of participating in household purchases and personal earnings (i.e., financial matters). This can be understood through the role of cousin marriage in solidifying economic and financial resources within the household. Financial decisions often involve considerations of economic resources, stability, and future planning. In households with cousin marriages, there may be a greater emphasis on maintaining economic stability and managing resources in ways that align with family expectations. This could lead to women being sidelined or having less influence in financial decision-making processes. The second hypothesis of this study posits that cousin marriage serves as an institution that reinforces and perpetuates restrictive gender roles. The Pakistan Demographic Health Survey 2018-19 includes data on women's attitudes towards accepting spousal violence, which is captured through 5 variables (V744a-V744e). (V744a-V744e). By linking cousin marriages with higher acceptance or justification of spousal violence, I argue that these marriages may contribute to normalizing violence within marital relationships and perpetuating domestic abuse/intimate partner violence). This normalization is often grounded in beliefs that women should conform to specific behaviors and roles, and that deviation from these roles can justify or warrant punitive actions. A preliminary analysis of this data in Table 4 reveals that a larger percentage of women in cousin marriages justify acts of spousal violence under certain circumstances. For instance, they are more likely to agree that a husband is justified in hitting his wife if she violates culturally acceptable roles of a housewife, such as going out without informing her husband, burning the food, neglecting children, arguing with her husband, or refusing to have sex with him, compared to women in non-cousin marriages.

**Table 4: Married women's attitudes towards spousal violence by marriage type**

|  | Women in non-cousin marriages (n=6,656) | Women in cousin marriages (n=8412) | Percentage difference (P-value) |
|---|---|---|---|
|  | N (%) | N (%) |  |
| Participant believes hitting is justified if a wife: |  |  |  |
| Goes out without telling her husband | 6534 (34) | 8209 (40) | 6 (0.000) |
| Neglects the children | 6516 (29) | 8193 (33) | 4 (0.000) |
| Argues with her husband | 6543 (26) | 8210 (31) | 5 (0.000) |
| Burns the food | 6520 (18) | 8161 (21) | 3 (0.000) |
| Refuses to have intercourse with her husband | 6445 (29) | 8079 (34) | 5 (0.000) |

Notes: Each of the variables on the justification of spousal violence is coded as 1 if the participant believes the husband is justified in hitting his wife and 0 if she believes it is not justified. The number of respondents in the subsample is in the parenthesis. Column 4 shows the p-value for a test of proportion with the null hypothesis (under the chi-square test) of no difference across groups.

Previous literature has consistently linked women's attitudes toward spousal violence with their adherence to patriarchal gender norms, which prescribe women's submissive social and familial roles (Zaatut and Haj-Yahia, 2016; Tayyab et al., 2017). These attitudes may also reflect the indoctrination of ideals about rigid gender and spousal roles (for a summary review of relevant literature, see Haj-Yahia, 1998). Yount, Zureick-Brown, and Salem (2014) used a case study from a village in Egypt to demonstrate that recent and chronic exposure to spousal violence was associated with higher domestic workload among women. They suggest that women may increase their involvement in household chores to conform to local gender norms.

The results in Table 4 indicate that approximately 40% of women in cousin marriages justify spousal violence if the wife goes out without informing her husband, compared to around 34% of women in non-cousin marriages— a difference of 6 percentage points or nearly 18%. Women in cousin marriages are also 19% more likely to justify spousal violence in cases where the wife argues with her husband, compared to women in non-cousin marriages.

Findings from Table 4 also suggest that, on average, women in both non-cousin and cousin marriages tend to justify spousal violence more frequently if the wife goes out without informing her husband, compared to scenarios such as neglecting children, arguing with her husband, burning food, or refusing intercourse with her husband.

Attitudes towards violence against wives vary significantly across countries and are influenced by various societal factors. For example, Rani and Banu (2009) conducted a comparative analysis using demographic and health surveys from seven countries (Armenia, Bangladesh, Cambodia, India, Kazakhstan, Nepal, and Turkey) between 1998 and 2001. The surveys presented respondents with scenarios involving transgressions from traditional gender norms—such as burning food, arguing with a husband, going out without informing him, neglecting children, or refusing sexual relations—and found that acceptance of wife-beating as punishment was reported by at least 30% of both men and women in all these countries. India showed the highest rates at 57%, followed by Turkey at 49%, despite both countries having comprehensive laws addressing domestic violence. Acceptance levels in Nepal (29%) were lower than those in Armenia (35%) and Kazakhstan (32.3%), where female literacy rates were comparatively higher. In Jordan, both men and women were found to be equally likely to justify wife-beating (60% of men and 62% of women) (Khawaja et al., 2008).

Similarly, Tran, Nguyen, and Fisher (2016) analyzed UNICEF Multiple Indicator Cluster Surveys across 39 countries to explore attitudes toward wife-beating or intimate partner violence (IPV) against women. They identified wide variation, with proportions of women endorsing the acceptance of 'wife-beating' ranging from 2.0% in Argentina to 90.2% in Afghanistan. Among men, acceptance rates varied from 5.0% in Belarus to 74.5% in the Central African Republic. This acceptance was most prevalent in Africa and South Asia, while least common in Central and Eastern Europe, Latin America, and the Caribbean. Generally, acceptance of 'wife-beating' was more prevalent among individuals in disadvantaged circumstances, including those in the lowest household wealth quintile, residing in rural areas, and having limited formal education.

Studies indicate that social norms and beliefs about traditional gender roles can be intergenerationally transmitted (Schmitz and Spiess, 2022). In this context, cousin marriage may act as a catalyst in perpetuating these norms. To explore the relationship between cousin marriage and married women's attitudes towards spousal violence, a logit regression model is estimated in Table C (see the appendix). Drawing on a conceptual framework similar to that employed by Rani and Banu (2009) and relevant literature, predictor variables are selected based on their potential influence on women's attitudes toward justifying hitting or beating a wife if she deviates from the traditional gender roles.

The findings indicate that being in a cousin marriage correlates with a higher likelihood of justifying or accepting spousal violence, even after accounting for other factors. These findings can be understood through several potential sociocultural and economic dynamics. In environments where spousal violence is prevalent and normalized, women may internalize these behaviors as acceptable or even necessary in certain circumstances. They may view violence as a legitimate response to perceived transgressions or breaches of marital expectations, such as disobedience or failure to fulfill domestic duties. Some women may also experience psychological manipulation or coercion from their spouses, leading to a distorted perception of the violence they endure. This can include feelings of guilt, shame, or self-blame for deviation from traditional gender roles which can further reinforce their justification of spousal violence as deserved or justified (see Dhanaraj and Mahambare, 2022).

Overall, these findings provide valuable insights into the literature on spousal violence, conformity to prevalent gender norms, and women's paid work. Cousin marriages may exacerbate and perpetuate spousal violence against women, as various studies suggest an intergenerational transmission of attitudes towards spousal violence, with women exposed to inter-parental violence more likely to justify and experience spousal violence (for example, see Dasgupta, 2019). In this regard, cousin marriage can be viewed as an institutional structure that reinforces restrictive gender norms and imposes penalties for deviating from conventional gender roles, which can directly impact women's labor supply patterns.

**b) Cousin marriage and women's paid work: Regression results and discussion**

To examine the association between cousin marriage and women's labor supply patterns (participation in paid work, work at home/way, occupation type—service-related/agricultural or manual), logit regression models are estimated in Tables 5 and 6. In Table 5, the outcome variable is dichotomous and indicates whether a woman participates in work for cash or in-kind remuneration after marriage (coded as 1) or not (coded as 0). In Table 6, two other binary dependent variables are tested under alternate estimation models: work at home (coded as 1 if a woman works at home, 0 if she works away from home) and occupation type (coded as 1 if a woman engages in service-related occupations, 0 if she is employed in agricultural and manual occupations). The explanatory variable of interest is cousin marriage, indicating marriage to a first or second cousin. The regression model also includes control variables for individual and household characteristics (age, education, husband's education, husband's work status, number of children, gender of the household head, household wealth, ethnicity, and residence in rural/urban areas). Moreover, region-level dummies are included to account for unobserved regional differences. Table 5 presents the log odds (i.e., logits)[22] for the association between cousin marriage and women's paid work. A negative logit indicates that as cousin marriage increases, the log odds of women's paid work decrease. Overall, after accounting for freedom of mobility (coded as 1 if a woman participates in planning family visits and 0 if she does not) and other personal and household attributes, women who are married to their cousins appear significantly less likely to be in paid work than women in non-cousin marriages. The average marginal effects from the logistic regression in Stata (see Table D in the appendix) show that women in cousin marriages have approximately a 2-percentage point lower probability of being in paid work compared to women who are not married to their cousins. This result carries substantial practical significance given Pakistan's low female labor force participation rate, which is only 21%, well below the South Asian average rate of 26% (World Bank Data, 2022).

Consistent with the previous literature, married women's age and education are positively associated with their engagement in paid work. Conversely, the husband's education level, number of children, and household wealth negatively influence married women's paid work (see Khan and Khan, 2009; Faridi and Rashid, 2014). The postestimation average marginal effects indicate that as household wealth increases, there is a decreasing probability of women being in paid work. The magnitude of this effect becomes more pronounced with higher levels of wealth.

---

[22] There is a direct relationship between the logit coefficients produced and the odds ratios. A logit is defined as the natural logarithm(log) of the odds: logit (p)= log(p/1-p)

**Table 5: Cousin marriage and women's paid work in Pakistan**
**Dependent variable: Paid work (1= work for cash or in-kind)**

| VARIABLES | (1)<br>Logit | (2)<br>Logit |
|---|---|---|
| Cousin marriage | -0.165*** | -0.158*** |
|  | (0.051) | (0.0512) |
| Age | 0.043*** | 0.038*** |
|  | (0.003) | (0.003) |
| Education | 0.108*** | 0.104*** |
|  | (0.007) | (0.007) |
| Freedom of mobility |  | 0.412*** |
|  |  | (0.053) |
| Education of husband | -0.036*** | -0.037*** |
|  | (0.006) | (0.006) |
| Husband's work status | -0.022 | -0.053 |
|  | (0.109) | (0.110) |
| Number of children | -0.060*** | -0.050*** |
|  | (0.019) | (0.019) |
| Gender of household head | 0.197** | 0.238*** |
|  | (0.088) | (0.088) |
| Rural | -0.048 | -0.030 |
|  | (0.058) | (0.059) |
| Wealth Index(ref.= poorest): |  |  |
| Poorer | -0.389*** | -0.398*** |
|  | (0.079) | (0.080) |
| Middle | -0.477*** | -0.494*** |
|  | (0.088) | (0.088) |
| Richer | -0.776*** | -0.790*** |
|  | (0.101) | (0.101) |
| Richest | -1.054*** | -1.076*** |
|  | (0.116) | (0.116) |
| Constant | -2.94*** | -3.04*** |
|  | (0.222) | (0.223) |
| Observations | 14,462 | 14,462 |

Note: The dependent variable is a binary variable, taking 1 if a woman works for cash or in-kind and 0 if she does not. Standard errors are provided in parentheses: * $p<0.05$, ** $p<0.01$, *** $p<0.001$. The models controlled for the effects of region and ethnicity.

The regression results in Table 6 examine how cousin marriage influences the nature of women's employment in Pakistan, focusing on their likelihood of engaging in work at home and their occupation type.

### Table 6: Cousin marriage and the quality of women's work in Pakistan

|  | (1) Work at home (Logit) | (2) Occupation type (Logit) |
| --- | --- | --- |
| Cousin Marriage | 0.302** | -0.187 |
|  | (0.103) | (0.121) |
| Age | -0.029*** | 0.031*** |
|  | (0.006) | (0.008) |
| Education | -0.130*** | 0.197*** |
|  | (0.013) | (0.016) |
| Education of husband | 0.041*** | 0.014 |
|  | (0.012) | (0.015) |
| Husband's work status | -0.116 | -0.453 |
|  | (0.205) | (0.242) |
| Number of Children | -0.008 | -0.097 |
|  | (0.038) | (0.051) |
| Gender of household head | -0.330 | 0.796*** |
|  | (0.172) | (0.215) |
| Rural | -0.086 | -0.773*** |
|  | (0.118) | (0.137) |
| Wealth Index (ref. =poorest): |  |  |
| Poorer | 0.815*** | 0.370 |
|  | (0.147) | (0.204) |
| Middle | 1.003*** | 0.087 |
|  | (0.169) | (0.221) |
| Richer | 0.976*** | 0.139 |
|  | (0.205) | (0.252) |
| Richest | 0.512** | 0.677* |
|  | (0.239) | (0.289) |
| Constant | 1.614*** | -3.08*** |
|  | (0.437) | (0.544) |
| Observations | 2,111 | 2,105 |

Note: The dependent variable in column 1 is binary, taking 1 if a woman works at home (for cash or in-kind) and 0 if she works away from home. In column 2, the dependent variable takes 1 if a woman works in service-related occupations and 0 if she performs agricultural or manual labor. Standard errors are provided in parentheses: * $p < 0.05$, ** $p < 0.01$, *** $p < 0.001$. The models controlled for the effects of region and ethnicity.

In column 1 of table 6, the dependent variable captures women's work at home or away. The results show that cousin marriage is associated with positive log odds of women working at home. This suggests that women in cousin marriages are more likely to engage in home-based work, possibly due to stricter social constraints and a higher burden of domestic responsibilities. Results in column 2, however, show that cousin marriage does not significantly influence the likelihood of women working in service-related occupations versus agricultural or manual labor. Among other factors, age and education levels show significant associations with both work at home and occupation type. Older age is negatively associated with the likelihood of working at home, while higher education levels are negatively associated with both working at home and working in service-related occupations. The likelihood of working at home significantly decreases with a married woman's age and education (number of years of schooling) and increases with her husband's education and household wealth.

The average marginal effects (see Table D in the appendix) show that higher wealth categories are associated with a higher probability of working at home compared to lower wealth categories. Women in the middle and richer wealth quintiles have on average, a 21% and 20% higher probability of working at home compared to the respective reference groups. The finding that wealthier women are more likely to engage in home-based work may reflect different socio-economic roles and complex gender norms that interact with higher economic status. These results also correspond well with the findings from Bangladesh that women prefer informal work within the home over options outside the home due to cultural norms that stigmatize women's work outside the home and their domestic burden of care. Women might prefer working at kin-based enterprises as they are not necessarily subjected to purdah (veil) (Heintz et al., 2018). In a previous study, nearly 40% of women in Pakistan reported that they don't pursue paid work as family members do not allow them to work outside the home (Tanaka and Muzones, 2016). In Pakistan, where 50% of female employment comprises contributing family workers[23] engaged in predominantly informal and unpaid roles (World Bank, 2022)[24], these findings highlight significant socio-economic dynamics. If norms of purdah and other gender norms are strictly enforced on women of higher social status, this may discourage their paid work outside the home. The results also align with similar studies in South Asia that find that domestic work at home is perceived as 'status-producing' and is more prevalent in households with higher incomes (Naidu, 2016). Status considerations may lead women to avoid work involving contact with non-family males. Additionally, activities categorized as 'status production' demand significant time and effort from married women. Less affluent women predominantly rely on paid work for income, often outside the home due to limited assets and resources. However, as incomes rise, concerns about social status become more prominent. This trend may prompt married women to gradually reduce their engagement in market work outside the home. Ultimately, at higher levels of affluence, they may cease market work entirely to focus on activities that enhance their social standing (Eswaran et al., 2013).

[23] Based on the International Labour Organization (ILO) definition, contributing family work involves market-oriented tasks for family members, such as cash, in-kind remuneration, or unpaid labor. Compensation is irregular and may include non-monetary benefits.

[24] Compared to other Asian countries with similar social and cultural norms, Pakistan has the highest share of female contributing family workers. India and Bangladesh, the share of contributing family work is around 29% and 27%, respectively.

It's important to note that while these findings offer insights into the relationship between marriage dynamics and the quality of women's work, caution is warranted due to potential endogeneity concerns. The causation between cousin marriage and women's paid work could operate in reverse (i.e., women already engaged in paid work may be less likely to marry their cousins), or there may be issues of simultaneity where cousin marriage and paid work mutually influence each other simultaneously. Therefore, it's important to interpret the results in Tables 5 and 6 as indicative of a correlation rather than implying causation between cousin marriage and the type of women's paid work. Nevertheless, these findings provide preliminary insights into the role of cousin marriage in various aspects of women's lives, including household decision-making, attitudes toward spousal violence, and patterns of labor supply. They underscore the barriers and constraints arising from cultural norms and familial expectations that influence women's decisions regarding paid work within a unique cultural and religious context. These insights highlight the challenges women face in balancing traditional domestic roles with economic participation.

### c) Sensitivity checks

To evaluate the sensitivity of the association between cousin marriage and women's work, robustness checks are implemented. Firstly, I find that the results in Tables 5 and 6 remain robust across different specifications and estimation methods. In Table B of the appendix, regression results employing Ordinary Least Squares (OLS) are provided. The sign and size of the coefficient on cousin marriage remained robust to differences in specifications and estimation methods.[25]

Secondly, it is informative to explore how the closeness of blood relation between spouses may influence women's engagement in paid work. Given the type of marriage, the nature of closeness between affinal and natal families differs in terms of emotional social, and economic factors. For example, family norms, solidarity, and moral values are considered more stable in first-cousin marriages than among distant relatives (Bittles, 2008, Otto et al., 2020). In this vein, societies, where first-cousin marriages are widely prevalent, tend to have more rigid patriarchal norms (Edlund, 2018). Thus, women married to first cousins may face more constraints (e.g., restrictions on mobility and household decision-making and other restrictive norms) than their counterparts in other types of marriages. As the previous regression model focused specifically on marriages to first and second cousins, it's also important to draw a comparison of marriages to kin and non-kin. This approach will allow a comparison of marriage to a first cousin, second cousin, and other relatives with the reference category as marriage to a non-relative.

The Pakistan Demographic and Health Survey 2018-19 included data on different types of marriages, such as those between blood relatives and non-relatives. Participants were asked whether they share a blood relation with their husbands, with responses recorded as "yes" or "no". This variable helps in investigating whether the closeness of a marital union might have varying effects on women's paid work. The marginal effects under a logistic regression analysis are presented in Table 7. Average marginal effects under logistic regression show that marriage to a blood relative is associated with a lower probability of women's participation in paid work. Including alternate explanatory variables for consanguineous marriage (marriage to a first cousin, second cousin, or a distant relative) does not significantly change the results. In column 2, the regression model further tests for the closeness of blood relation with the husband. The independent variable of interest is "marriage to a paternal first cousin" which takes 1 if a woman is married to a first cousin on the father's side and 0 if she is married to other cousins or non-relatives. The results are consistent with previous estimations with minimal differences.

---

[25] Similar findings were observed across various indicators of autonomy, with one notable exception: when controlling for self-earnings, the coefficient associated with cousin marriage becomes statistically insignificant.

**Table 7: Marriage type and women's paid work: marginal effects**

| VARIABLES | (1) Paid work | (2) Paid work |
|---|---|---|
| **Marriage to a relative** | -0.018*** | |
| | (0.006) | |
| **Marriage paternal first cousin** | | -0.020*** |
| | | (0.007) |
| Age | 0.005*** | 0.005*** |
| | (0.000) | (0.000) |
| Education | 0.012*** | 0.012*** |
| | (0.001) | (0.001) |
| Education of Husband | -0.004*** | -0.004*** |
| | (0.001) | (0.001) |
| Husband's work status | -0.003 | -0.003 |
| | (0.013) | (0.013) |
| Number of Children | -0.007** | -0.007** |
| | (0.002) | (0.002) |
| Gender of household head | 0.024** | 0.023** |
| | (0.010) | (0.010) |
| Rural | -0.005 | -0.006 |
| | (0.007) | (0.007) |
| Wealth Index (ref. =poorest): | | |
| Poorer | -0.045*** | -0.054*** |
| | (0.009) | (0.011) |
| Middle | -0.052*** | -0.066*** |
| | (0.010) | (0.012) |
| Richer | -0.082*** | -0.099*** |
| | (0.011) | (0.013) |
| Richest | -0.113*** | -0.124*** |
| | (0.013) | (0.014) |
| Observations | 14,459 | 14,462 |

Notes: The dependent variable is paid work. The independent variable of interest is marriage to a relative (equals 1 if a woman is married to a blood relative, 0= if there is no blood relation). Standard errors in parentheses: *** $p<0.01$, ** $p<0.05$, * $p<0.1$. The model controlled for the effects of region and ethnicity.

This study also tested for overall average marginal effects of cousin marriage (1=marriage to first and second cousins, 0= otherwise) on women's paid work after accounting for freedom of mobility and additional attributes. Table D in the Appendix presents consistent results. The estimates (columns 1 and 2) suggest that women in cousin marriages have approximately a 2-percentage point lower probability of being in paid work compared to women who are not in cousin marriages, holding all other variables constant. Column 3 in Table D examines the difference in the probability of working at home between the two groups (cousin marriages vs. non-cousin marriages) after controlling for freedom of mobility and other attributes. The results suggest that women in cousin marriages are, on average, 6.5 percentage points more likely to work at home compared to women not in cousin marriages. Consistent with the previous results, as household wealth increases from poorer to richer categories, women are less likely to engage in paid work and more likely to work at home.

Overall, while the correlation between cousin marriage and women's paid work appears weak, its association with the quality of women's work, particularly in terms of contributing to family work, and women's attitudes towards traditional gender roles remains worthwhile and deserves further exploration.

## V. CONCLUSION

Various studies caution against narrowly focusing on women's bargaining power within households and their labor supply patterns without considering the broader contexts of kinship networks and communities. For example, Folbre (1994) argues that larger structures of constraint—assets, rules, norms, and preferences that define the boundaries of what individuals want and how they can pursue it—play a significant role in shaping economic decisions for both men and women.

This study uses data from 15,068 ever-married women aged 15-49 from the Pakistan Demographic Health Survey, 2017-18, to analyze the role of cousin marriage, which reinforces kinship networks and norms, in women's status in the household, attitudes towards spousal violence, and participation in paid work. Marriages between first and second cousins constitute more than half of all marital unions in Pakistan. Despite the rigid patriarchal gender roles associated with cousin marriage, it may be favored due to its perceived social and economic benefits over other forms of marital unions. These include stronger kinship ties, a lower likelihood of divorce, and the potential to negotiate lower dowries and bride prices. Interestingly, approximately 83% of women in cousin marriages and nearly 77% in other marriages report being involved in selecting their husbands.

This study offers several key findings. Firstly, cousin marriage does not improve women's autonomy or bargaining power in the household and appears to have a weak negative correlation with women's participation in financial decisions. Secondly, women in cousin marriages tend to hold more restrictive patriarchal attitudes towards gender roles in the household, justifying spousal violence against wives who do not conform to traditional gender roles. Finally, cousin marriage shows a modest, negative correlation with women's participation in paid work.

The paper also explores how cousin marriage influences the qualitative nature of women's work. Results show that women in cousin marriages are more likely to work at home compared to those in non-cousin marriages. This observation holds significant practical implications, particularly given that a substantial majority of female employment in Pakistan (54%) consists of contributing family workers, unpaid and informal. It is possible that women in cousin marriages face greater constraints due to expectations regarding their contributions to home production and their adherence to patriarchal gender and family ideals.

This study contributes to the existing literature by highlighting the association between cousin marriage and various constraints women may face in pursuing economic opportunities. However, due to the endogenous nature of marriage decisions, causal links cannot be asserted. Nevertheless, these findings provide valuable insights into how marriage dynamics and patriarchal gender roles are interconnected, influencing women's participation in paid work. Cousin marriage may act as an institution that reinforces stringent gender norms, limiting women's participation in the public sphere, especially in societies where such marriages are prevalent.

**Acknowledgments:** I extend my sincere gratitude to the participants of the 28th International Association of Feminist Economics Conference and the 2023 International Behavioral Public Policy Conference at the University of North Carolina at Chapel Hill, as well as the esteemed faculty at the University of Massachusetts Amherst and the University of Washington (Tacoma), for their invaluable feedback and suggestions. I also appreciate the comments, suggestions, and questions provided by the two anonymous reviewers, whose thoughtful insights significantly enriched the discussion and analysis presented in this paper. Finally, I am grateful for the support under the US Fulbright Program and from the Political Economy Research Institute at the University of Massachusetts Amherst.

**Data Availability:** The dataset, description of the tools, and other supplementary material are available upon request.

**Appendix**

**Table A: Cousin marriage and women's status in the household**

| VARIABLES | (1) Visits to family | (2) Healthcare | (3) Household purchases | (4) Husband's earnings | (5) Own earning |
|---|---|---|---|---|---|
| Cousin marriage | -0.06 | -0.053 | -0.123** | -0.060 | -0.152** |
| | (0.038) | (0.037) | (0.038) | (0.038) | (0.056) |
| Age | 0.058*** | 0.045*** | 0.060*** | 0.044*** | 0.037*** |
| | (0.002) | (0.002) | (0.002) | (0.002) | (0.003) |
| Education | 0.044*** | 0.05*** | 0.046*** | 0.036*** | 0.115*** |
| | (0.005) | (0.005) | (0.005) | (0.004) | (0.007) |
| Education of husband | 0.007 | 0.004 | 0.000 | -0.000 | 0.037*** |
| | (0.004) | (0.001) | (0.004) | (0.004) | (0.006) |
| Husband's work status | 0.375*** | 0.239*** | 0.365*** | 1.517*** | -0.115 |
| | (0.083) | (0.080) | (0.085) | (0.103) | (0.116) |
| Number of children | -0.134*** | -0.111*** | -0.14*** | -0.133*** | -0.11*** |
| | (0.0135) | (0.013) | (0.014) | (0.013) | (0.022) |
| Gender of household head | -0.565*** | -0.742*** | -0.681*** | -0.798*** | 0.053 |
| | (0.063) | (0.063) | (0.062) | (0.062) | (0.094) |
| Constant | -1.49** | -0.168** | -1.88*** | -2.27*** | -3.00*** |
| | (0.164) | (0.042) | (0.165) | (0.174) | (0.241) |
| Observations | 14,469 | 14,469 | 14,469 | 14,469 | 14,469 |

Note: Regression estimates under the logit models are presented here. Each of the decision-making variables is coded as 1 if a woman alone or jointly with her husband is consulted in the decision-making process; 0 if she is not involved in the decisions (husband alone or someone else is involved). Standard errors are provided in parentheses: * $p<0.05$, ** $p<0.01$, *** $p<0.001$. The models controlled for household wealth, residence in rural/urban areas, and effects of region/ethnicity.

**Table B: The relationship between cousin marriage and women's paid work in Pakistan**

|  | (1)<br>Paid work | (2)<br>Work at home | (3)<br>Occupation type |
|---|---|---|---|
| Cousin Marriage | -0.019*** | 0.059*** | -0.022 |
|  | (0.006) | (0.022) | (0.017) |
| Age | 0.005*** | -0.007*** | 0.004*** |
|  | (0.000) | (0.0012) | (0.001) |
| Education | 0.012*** | -0.028*** | 0.036*** |
|  | (0.001) | (0.0028) | (0.002) |
| Education of husband | -0.003*** | 0.008*** | 0.002 |
|  | (0.001) | (0.003) | (0.002) |
| Husband's work status | -0.01 | 0.0019 | -0.061* |
|  | (0.0124) | (0.045) | (0.036) |
| Number of Children | -0.004* | -0.002 | -0.014** |
|  | (0.002) | (0.008) | (0.007) |
| Gender of household head | 0.0261*** | -0.066* | 0.104*** |
|  | (0.010) | (0.037) | (0.03) |
| Wealth Index (ref. =poorest): |  |  |  |
| Poorer | -0.05*** | 0.208*** | 0.008 |
|  | (0.010) | (0.033) | (0.026) |
| Middle | -0.071*** | 0.252*** | -0.029 |
|  | (0.0104) | (0.037) | (0.03) |
| Richer | -0.106*** | 0.248*** | -0.004 |
|  | (0.012) | (0.045) | (0.036) |
| Richest | -0.145*** | 0.179*** | 0.062 |
|  | (0.013) | (0.053) | (0.042) |
| Rural | -0.0097 | -0.013 | -0.106*** |
|  | (0.007) | (0.026) | (0.021) |
| Constant | 0.0711*** | 0.748*** | 0.056 |
|  | (0.0265) | (0.099) | (0.079) |
| Observations | 14,462 | 2,111 | 2,105 |
| R-squared | 0.093 | 0.154 | 0.426 |

Notes: OLS coefficients with robust standard errors are in parentheses: *** *p<0.01, ** p<0.05, * p<0.1* . The models controlled for the effects of region/ethnicity.

**Table C: Cousin marriage and married women's attitudes towards spousal violence**
**Dependent variable: Accept Spousal violence**

|  | (1) | (2) |
|---|---|---|
|  | Logit | Logit |
| Cousin marriage | 0.104** | 0.131** |
|  | (0.050) | (0.051) |
| Age | -0.003 | -0.002 |
|  | (0.003) | (0.003) |
| Education | -0.09*** |  |
|  | (0.007) |  |
| Paid work |  | 0.209** |
|  |  | (0.071) |
| Education of husband | -0.016** | -0.036*** |
|  | (0.005) | (0.005) |
| Husband's work status | -0.093 | -0.101 |
|  | (0.097) | (0.097) |
| Number of Children | 0.073*** | 0.083*** |
|  | (0.015) | (0.015) |
| Gender of household head | -0.026 | -0.014 |
|  | (0.085) | (0.085) |
| Rural | 0.117 | 0.135** |
|  | (0.060) | (0.060) |
| Wealth Index (ref. =poorest): |  |  |
| Poorer | -0.345*** | -0.374*** |
|  | (0.065) | ((0.065) |
| Middle | -0.585*** | -0.708*** |
|  | (0.080) | (0.080) |
| Richer | -0.739*** | -0.979*** |
|  | (0.098) | (0.096) |
| Richest | -1.014*** | -1.48*** |
|  | (0.124) | (0.119) |
| Constant | -1.09*** | -1.44*** |
|  | (0.196) | (0.196) |
| Observations | 14,294 | 14,294 |

Note: Regression estimates under the logit models are presented here. The dependent variable is a binary variable capturing women's perceptions of spousal violence. Here, the variable "accept spousal violence" is coded as 1 if a woman believes that it is justified for a husband to hit his wife for any of the following actions: going out without informing her husband, burning the food, neglecting children, arguing with her husband, or refusing to have sex with him. It is coded as 0 otherwise. Standard errors are provided in parentheses: * $p<0.05$, ** $p<0.01$, *** $p<0.001$. The models controlled for the effects of region/ethnicity.

**Table D: Cousin marriage and women's labor supply patterns: average marginal effects from logit models**

| VARIABLES | (1)<br>Paid work | (2)<br>Paid work | (3)<br>Work at home |
|---|---|---|---|
| Cousin marriage | -0.019*** | -0.018** | 0.065** |
| | (0.006) | (0.006) | (0.022) |
| Age | 0.005*** | 0.004*** | -0.006*** |
| | (0.000) | (0.000) | (0.001) |
| Education | 0.012*** | 0.012*** | -0.028*** |
| | (0.001) | (0.001) | (0.003) |
| Freedom of mobility | | 0.047*** | 0.021 |
| | | (0.006) | (0.022) |
| Education of husband | -0.004*** | -0.004*** | -0.009*** |
| | (0.001) | (0.001) | (0.003) |
| Husband's work status | -0.003 | -0.006 | 0.001 |
| | (0.013) | (0.013) | (0.045) |
| Number of children | -0.007** | -0.006** | 0.002 |
| | (0.002) | (0.002) | (0.008) |
| Gender of household head | 0.023* | 0.027** | -0.074* |
| | (0.01) | (0.01) | (0.037) |
| Rural | -0.006 | -0.003 | 0.021 |
| | (0.007) | (0.007) | (0.025) |
| Wealth Index(ref.= poorest): | | | |
| Poorer | -0.054*** | -0.055*** | 0.173*** |
| | (0.011) | (0.080) | (0.032) |
| Middle | -0.065*** | -0.067*** | 0.221*** |
| | (0.012) | (0.012) | (0.036) |
| Richer | -0.098*** | -0.10*** | 0.211*** |
| | (0.013) | (0.013) | (0.044) |
| Richest | -0.124*** | -0.126*** | 0.118* |
| | (0.014) | (0.014) | (0.051) |
| Observations | 14,462 | 14,462 | 2,111 |

Note: The table shows average marginal effects under the logit model. For brevity, estimates for the control variables related to region and ethnicity are not included. Standard errors are provided in parentheses: * $p<0.05$, ** $p<0.01$, *** $p<0.001$.